\documentclass[aps,prl,amssymb,twocolumn,showpacs]{revtex4}
\usepackage{graphicx}

\newcommand{\bea}{\begin{eqnarray}}

\newcommand{\eea}{\end{eqnarray}}

\newcommand{\beq}{\begin{equation}}

\newcommand{\eeq}{\end{equation}}

\newlength{\textwidthm}

\begin{document}

\def \tr{{\mbox{tr~}}}
\def \ra{{\rightarrow}}
\def \ua{{\uparrow}}
\def \da{{\downarrow}}
\def \be{\begin{equation}}
\def \ee{\end{equation}}
\def \ba{\begin{array}}
\def \ea{\end{array}}
\def \bea{\begin{eqnarray}}
\def \eea{\end{eqnarray}}
\def \nn{\nonumber}
\def \l{\left}
\def \r{\right}
\def \half{{1\over 2}}
\def \etal{{\it {et al}}}
\def \cH{{\cal{H}}}
\def \cM{{\cal{M}}}
\def \cN{{\cal{N}}}
\def \cQ{{\cal Q}}
\def \cI{{\cal I}}
\def \cV{{\cal V}}
\def \cG{{\cal G}}
\def \cF{{\cal F}}
\def \cZ{{\cal Z}}
\def \bS{{\bf S}}
\def \bI{{\bf I}}
\def \bL{{\bf L}}
\def \bG{{\bf G}}
\def \bQ{{\bf Q}}
\def \bK{{\bf K}}
\def \bR{{\bf R}}
\def \br{{\bf r}}
\def \bu{{\bf u}}
\def \bq{{\bf q}}
\def \bk{{\bf k}}
\def \bz{{\bf z}}
\def \bx{{\bf x}}
\def \bpsi{{\bar{\psi}}}
\def \tJ{{\tilde{J}}}
\def \W{{\Omega}}
\def \e{{\epsilon}}
\def \lam{{\lambda}}
\def \L{{\Lambda}}
\def \a{{\alpha}}
\def \t{{\theta}}
\def \b{{\beta}}
\def \g{{\gamma}}
\def \D{{\Delta}}
\def \d{{\delta}}
\def \w{{\omega}}
\def \s{{\sigma}}
\def \f{{\varphi}}
\def \x{{\chi}}
\def \e{{\epsilon}}
\def \h{{\eta}}
\def \G{{\Gamma}}
\def \z{{\zeta}}
\def \hatt{{\hat{\t}}}
\def \hn{{\bar{n}}}
\def \vk{{{\bf k}}}
\def \vq{{{\bf q}}}
\def \gk{{\g_{\vk}}}
\def \nd{{^{\vphantom{\dagger}}}}
\def \yd{^\dagger}
\def \av#1{{\langle#1\rangle}}
\def \ket#1{{\,|\,#1\,\rangle\,}}
\def \bra#1{{\,\langle\,#1\,|\,}}
\def \braket#1#2{{\,\langle\,#1\,|\,#2\,\rangle\,}}

\title{Competing orders in two-dimensional Bose-Fermi Mixtures}

\author{L.~Mathey$^1$, S.~-W.~Tsai$^2$, and A.~H.~Castro~Neto$^3$}

\affiliation{$^1$Physics Department, Harvard University, Cambridge, MA 02138 \\
$^2$Department of Physics, University of California, Riverside, CA 92521 \\
$^3$Department of Physics, Boston University, 590 Commonwealth Ave., Boston, MA 02215
}

\date{\today}

\begin{abstract}
Using a functional renormalization group approach we study the zero
 temperature phase diagram of two-dimensional Bose-Fermi mixtures of
 ultra-cold atoms in optical lattices, in the limit when the velocity of
 bosonic condensate fluctuations are much larger than the Fermi velocity.
 For spin-$1/2$ fermions we obtain a phase diagram, which
 shows a competition of pairing phases of various orbital
 symmetry ($s$, $p$, and $d$) and antiferromagnetic order. We
 determine the value of the gaps of various phases close to half-filling, and identify subdominant orders as well as short-range fluctuations
 from the RG flow. For spinless fermions we find that $p$-wave pairing dominates the phase diagram. 
\end{abstract}

\pacs{03.75.Hh,03.75.Mn,05.10.Cc}

\maketitle

Since the realization of the Mott insulator (MI) transition \cite{greiner} in ultra-cold atom systems, there has been remarkable progress in 'engineering' many-body states in a well-defined and tunable environment, due to the advances in trapping and manipulating ultra-cold atoms in optical lattices \cite{stoeferle, mandel, koehl}. Important achievements include the creation of fermionic superfluids \cite{BECBCS}, the realization of the Tonks-Girardeau gas \cite{Tonks} and of Luttinger liquids \cite{Luttinger}, and the observation of noise correlations \cite{noise}. One of the subjects of intense study recently has been mixtures of
ultra-cold bosonic and fermionic atoms. Experiments include 
 the condensation of molecules in mixtures of fermionic atoms 
\cite{BECBCS}, and the observation of instabilities in Bose-Fermi mixtures
(BFM) \cite{inst}.  Many other intriguing many-body phenomena have been
proposed, such as the appearance of charge density wave order (CDW)
\cite{cdw}, de-mixing transition \cite{leonid}, the formation of composite particles \cite{composite}, and polaronic effects \cite{mathey}.
Mixtures of ultracold bosonic and fermionic atoms that are
 subjected to optical lattices, which confine the atoms to move
 on a 2D lattice, exhibit one of the most intriguing phenomena of condensed matter physics, the competition of orders in many-body systems, which is also  observed in a variety of materials, such as high-$T_c$ compounds \cite{harlingen} and Bechgaard salts \cite{bechgaard}. To capture the properties of the many-body state of these systems presents a significant theoretical challenge. 

In this paper we study these systems using a functional renormalization group
(RG) approach introduced in the context of the stability of the Fermi liquid
fixed point \cite{shankar}, and that has been extensively applied to
interacting electrons on lattices in the last few years
\cite{Zanchi_Schulz,RG}, and has recently been
extended to include retardation effects associated with the electron-phonon
interaction \cite{tsai,honer}. Contrary to mean-field and variational
approaches \cite{wang}, the RG  approach includes corrections to the full
four-point vertices in the flow equations, and therefore treats all types of
order in an unbiased way. In this way we obtain the phase diagram and the
values of the gaps of the different types of order. Furthermore, we can read
off the subdominant orders and the short distance fluctuations from the RG
analysis. 

The fermionic atoms in the BFM considered here are prepared 
either  as a mixture of two hyperfine states (which we treat in an isospin language),  to create spin-$1/2$ fermions, or in a single hyperfine state to create spinless fermions. Fermionic atoms  in different hyperfine states interact via short-range, i.e., on-site interaction, whereas spinless fermions are essentially non-interacting. Besides two-body contact interactions, density fluctuations in a condensate of bosonic atoms induce attractive finite-range interactions between fermions, with a length scale given by the coherence length of the condensate. The competition between these two types of interaction leads to the possibilities of many types of different instabilities and, hence, to a rich phase diagram, as we discuss here. In the laboratory, this system can be realized as a $^{40}$K-$^{87}$Rb mixture in an optical lattice created by Nd:YAG lasers. The interaction between the different atomic species can be manipulated by either tuning the system close to a Feshbach resonance, or by using more
than one optical lattice to trap the different types of atoms and to spatially shift these lattices with respect to each other. 
  
Ultracold atoms in optical lattices are very accurately described
 by a Hubbard model. In the following we write the model for the case of spin-$1/2$ fermions. The case of spinless fermions can be immediately obtained from this by ignoring one of the spin states.
For a mixture of one type of bosonic atom and two fermionic types that are SU(2) symmetric, the Hamiltonian is given by:
\bea\label{Ham}
H & = & -t_{f}  \sum_{\langle i j \rangle,s} f^\dagger_{i,s} 
f_{j,s} - t_{b}  \sum_{\langle i j \rangle} b^\dagger_{i} b_{j}
- \sum_i (\mu_{f} n_{f,i} + \mu_b n_{b,i})
\nonumber\\
&+ &  \sum_{i} \Big[U_{ff} n_{f, i,\uparrow}n_{f,i, \downarrow}
+ \frac{U_{bb}}{2} n_{b, i}n_{b,i} + U_{bf} n_{b, i} n_{f,i} \Big] \, ,
\eea
where $f^\dagger_{i,s}$ ($f_{i,s}$) creates (annihilates) a fermion at site $i$ with pseudo-spin $s$ ($s=\uparrow,\downarrow$), $b^\dagger_i$ ($b_i$) creates (annihilates) a boson at site $i$,
$n_{f,i}= \sum_s f^\dagger_{i,s} f_{i,s}$ ($n_{b,i} = b^\dagger_i b_i$) is the fermion (boson) number operator,  $t_f$ and $t_b$ are the fermionic and bosonic tunneling energies between neighboring sites, 
$\mu_f$ ($\mu_b$) is the chemical potential for fermions (bosons), 
 $U_{bb}$ is the repulsion energy between bosons on the same site, $U_{ff}$ is the interaction energy between the two species of fermions, and $U_{bf}$ is 
the interaction 
energy between bosons and fermions. In momentum space, this Hamiltonian is written as:
\bea\label{Ham_k}
H & = & \sum_{\bold{k}} \left\{ (\epsilon_{f, \bold{k}} -\mu_f) \sum_s f^\dagger_{\bold{k},s} f_{\bold{k},s} + (\epsilon_{b, \bold{k}} -\mu_b) 
b^\dagger_{\bold{k}} 
b_{\bold{k}} \right.
\nonumber
\\
&+&\!\!\!\left.
\frac{U_{ff}}{V} \rho_{f,\bold{k},\uparrow}\rho_{f,-\bold{k}, \downarrow}
 \!\!+\!\! \frac{U_{bb}}{2 V}  \rho_{b, \bold{k}}\rho_{b,-\bold{k}}  \!\!+\!\! \frac{U_{bf}}{V} \rho_{b, \bold{k}}
\rho_{f,-\bold{k}} \right\}  \, ,
\eea
where $\rho_{f,{\bf k}} = \sum_{{\bf q},s}  f^\dagger_{\bold{k}+{\bf q},s}
f_{\bold{q},s}$ ($\rho_{b,{\bf k}} = \sum_{{\bf q}}  b^\dagger_{\bold{k}+{\bf
 q}} b_{\bold{q}}$) is the fermion (boson) density operator,  $\epsilon_{b/f,
 \bold{k}}= - 2 t_{b/f} (\cos k_x + \cos k_y)$, is the bosonic/fermionic
 dispersion relation for free particles on a square lattice. 

\begin{figure}
\includegraphics[width=7cm]{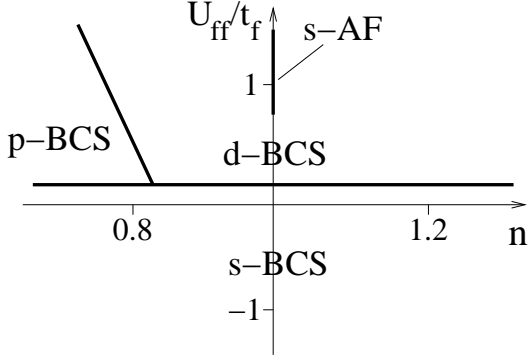}
\caption{\label{PD}
Phase diagram, interaction strength, $U_{ff}/t_f$, versus number of fermions per site, $n$, for a Fermi-Bose mixture in a square lattice
in 2D ($\tilde{V}/t_f = 2$, $\xi = 1$).}
\end{figure}

Throughout this paper we consider the limit of 
weakly interacting bosons that form a BEC. 
We  assume further that the velocity of the 
 condensate fluctuations
 is much larger than the Fermi velocity. These two 
assumptions  allow us to integrate out the bosonic degrees of 
freedom and use an instantaneous approximation, i.e. neglect retardation effects. 
In the limit of weakly interacting bosons,  the following approach is 
 well established:
We assume that the zero momentum bosonic mode is macroscopically
occupied, and we formally replace the corresponding operator
$b_0$ by a real number $b_0 \rightarrow \sqrt{N_0}$, where
$N_0$ is the number of condensed atoms.
After this replacement we keep all terms that
are quadratic in $b_\bold{k}$ (with $\bold{k}\neq0$), and
perform a Bogoliubov transformation, given by:
$b_\bold{k} = u_\bold{k} \beta_\bold{k} + v_\bold{k} \beta^\dagger_\bold{-k},$
to diagonalize the bosonic Hamiltonian.
The resulting eigenmodes $\beta_\bold{k}$ have a
 dispersion relation given by
$\omega_\bold{k} = \sqrt{\epsilon_{b,\bold{k}}(\epsilon_{b,\bold{k}} +2U_{bb}n_b)}$, 
 with the low-${\bf k}$ limit $\omega_\bold{k}\sim v_b|{\bf k}|$, 
with $v_b=\sqrt{2 t_b U_{bb} n_b}$.
The parameters $u_\bold{k}$ and $v_\bold{k}$ are given by:
$u_\bold{k}^2  =  (\omega_\bold{k} +\epsilon_{b,\bold{k}} + U_{bb} n_b)/(2 \omega_\bold{k})$ and 
$v_\bold{k}^2  =  (-\omega_\bold{k} + \epsilon_{b,\bold{k}} + U_{bb}n_b)/(2\omega_\bold{k})$.
The density fluctuations of the bosons are approximated by:
$\rho_{b,\bold{k}}\approx \sqrt{N_0}(u_\bold{k} +v_\bold{k})
(\beta_\bold{k} + \beta_{-\bold{k}}^\dagger)$, with $\bold{k}\neq 0$.
 The interaction between bosons and fermions is given by 
 $U_{bf}\sqrt{N_0}/V\sum_{\bold{k}}(u_\bold{k}+v_\bold{k})(\beta_\bold{k} 
+\beta_{-\bold{k}}^\dagger) \rho_{f,-\bold{k}}$.
 As a next step we integrate out the bosonic  modes and use an instantaneous  approximation, leading to the following effective
Hamiltonian:
\bea\label{Heff}
H_{{\rm eff.}} & = & \sum_{\bold{k} } \left\{ (\epsilon_{\bold{k}} -\mu_f) 
\sum_s f^\dagger_{\bold{k},s} 
f_{\bold{k},s} + \frac{U_{ff}}{V} \rho_{f,\bold{k},\uparrow}\rho_{f,-\bold{k}, \downarrow} \right.
\nonumber\\ 
&+&  \left. \frac{1}{2V} V_{{\rm ind.}, \bold{k}} \, \, \rho_{f,\bold{k},} \rho_{f,\bold{-k}} \right\}
\, ,
\eea
where the induced potential $V_{{\rm ind.}, \bold{k}}$ is given by:
$
V_{{\rm ind.}, \bold{k}}= - \tilde{V}/(1 + \xi^2 (4 - 2 \cos k_x -2 \cos k_y)) \, ,
$ 
with $\tilde{V}$ given by $\tilde{V}=U_{bf}^2/U_{bb}$, and $\xi$ is the
 healing length of the BEC and is given by  $\xi=\sqrt{t_b/2 n_b U_{bb}}$. Notice, that this approach is
 only valid when $v_b\gg v_f$, so that the fermion-fermion interaction mediated by the bosons can be considered as instantaneous. In
 the presence of retardation one cannot formally define a Hamiltonian
 formulation since the frequency dependence of the interaction appears
 explicitly. 
If that is the case, one has to consider the frequency dependence
 of the interaction explicitly as is done, 
for instance, in ref.~[\onlinecite{tsai}]. 
 Here, however, 
the 
system can be tuned to the 
 non-retarded limit, that is not the limit in
 most solid state systems.

Notice that (\ref{Heff}) describes the scattering of two fermions from momenta ${\bf k}_1$ and ${\bf k}_2$, that are scattered into momenta 
${\bf k}_3$ and ${\bf k}_4$. Momentum conservation at the interaction vertex
requires that ${\bf k}_4 = {\bf k}_1+{\bf k}_2 -{\bf k}_3$, and hence the
interaction vertex, $U(\bold{k}_1,\bold{k}_2,\bold{k}_3)$, depends on
three momenta \cite{remark_spin}. Its bare value can be written as:
\bea
U(\bold{k}_1,\bold{k}_2,\bold{k}_3) & = & U_{ff} + V_{{\rm ind.}, \bold{k}_1-\bold{k}_3} \, .
\label{Uk}
\eea

The RG is implemented by tracing out high energy degrees of freedom in a region between $\Lambda$ and $\Lambda + d\Lambda$, where 
$\Lambda$ is the energy cut-off of the problem. In this process, the vertex
$U$ is renormalized. At the initial value of the cut-off $\Lambda=\Lambda_0
\approx 8 t_f$, the value of $U$ is given by (\ref{Uk}), which is the initial
condition for the RG. The RG flow is obtained from a series of coupled
integral-differential equations \cite{Zanchi_Schulz} for all the different
interaction vertices $U(\bold{k}_1,\bold{k}_2,\bold{k}_3)$. From these, the
specific interaction channels, such as CDW, antiferromagnetic (AF),
and  
superconducting (BCS), can be identified:
%
%
\bea
\label{eq:cdw} V^{CDW} & = & 4 \, \, U_c(\bold{k}_1,\bold{k}_2, \bold{k}_1+\bold{Q}) \, ,
\\
\label{eq:af} V^{AF} & = & 4 \, \, U_\sigma(\bold{k}_1,\bold{k}_2, \bold{k}_1+\bold{Q}) \, ,
\\
\label{eq:bcs} V^{BCS} & = & U(\bold{k}_1,-\bold{k}_1, \bold{k}_2) \, ,
\eea
where we have use used: 
$U_c  =  (2-\hat{X})U/4$, $U_\sigma =  - \hat{X}U/4$ 
with $\hat{X}U(\bold{k}_1,\bold{k}_2,\bold{k}_3)=U(\bold{k}_2,\bold{k}_1,\bold{k}_3)$, and $\bold{Q}$ is the nesting vector, $\bold{Q}=(\pi,\pi)$. The RG equations read:
\begin{eqnarray}
&&\partial_{\ell} U_{\ell}({\bf k_1},{\bf k_2},{\bf k_3}) =
\nonumber\\
&-&\!\!\!\!\int_{p,\omega} \partial_{\ell}
[G_{\ell}({\bf p})G_{\ell}({\bf k})] U_{\ell}({\bf k_1},{\bf k_2},{\bf k})
U_{\ell}({\bf p},{\bf k},{\bf k_3})
\nonumber\\
&-&\!\!\!\!\int_{p,\omega} \partial_{\ell}
 [G_{\ell}({\bf p})G_{\ell}({\bf q_1})]
U_{\ell}({\bf p},{\bf k_2},{\bf q_1}) U_{\ell}({\bf k_1},{\bf q_1},{\bf k_3})
\nonumber\\
&-&\!\!\!\!\int_{p,\omega} \partial_{\ell}
 [G_{\ell}({\bf p})G_{\ell}({\bf q_2})] \left\{-\!2U_{\ell}({\bf k_1},{\bf p}, {\bf q_2})U_{\ell}({\bf q_2},{\bf k_2},{\bf k_3}) \right.
\nonumber\\
&\!+&\!\! \!\left. \!\!U_{\ell}(\!{\bf p},\!{\bf k_1},\!{\bf q_2}\!) U_{\ell}(\!{\bf q_2},\!{\bf k_2},\!{\bf k_3}\!)\! 
+\!  U_{\ell}(\!{\bf k_1},\!{\bf p},\!{\bf q_2}\!)
U_{\ell}(\!{\bf k_2},\!{\bf q_2},\!{\bf k_3}\!) \!\right\}\!\!,
\end{eqnarray}
where $\ell = \ln(\Lambda_0/\Lambda)$, ${\bf k} = {\bf k}_1+{\bf k}_2-{\bf p}$, ${\bf q}_1={\bf p}+{\bf k}_2-{\bf k}_3$, ${\bf q_2}={\bf p}+{\bf k}_1-{\bf k}_3$, and 
$G_{\ell}({\bf k}) = \Theta(|\xi_{{\bf k}}|-\Lambda)/(i \omega-\xi_{{\bf
    k}})$ with $\xi_{\bf k} = \epsilon_{f,{\bf k}} - \mu_f$.

\begin{figure}
\includegraphics[width=7.5cm]{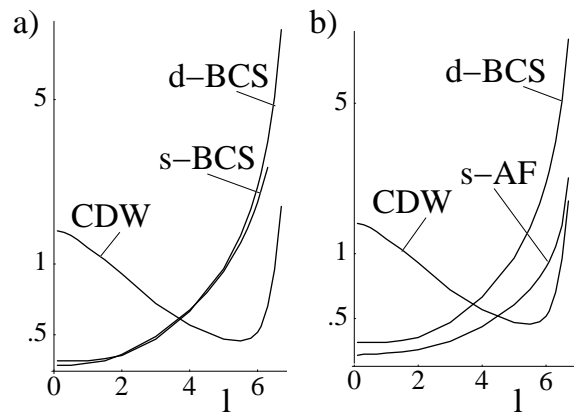}
\caption{\label{Flow_HF}
RG flow for the different effective interactions (in units of
$t_f$) as a function of the RG parameter $l$ ($\tilde{V}/t_f=3$ and $\xi=1$). a) $U_{ff}/t_F =0.5$; b) $U_{ff}/t_f=1.2$.}
\end{figure}

In our implementation we discretize the Fermi surface into $M=24$ patches,
and hence each of the interaction channels (\ref{eq:cdw}), (\ref{eq:af}),
(\ref{eq:bcs}) is represented by an $M\times M$ matrix. At each RG step, we
diagonalize each of these matrices. The channel with the largest
eigenvalue (with the caveat that a BCS-channel needs to be attractive to
drive a transition) corresponds to the dominant order. The elements of the
eigenvector are labeled by the discrete 
patch indices around the Fermi surface and the symmetry of the order
parameter is given by this angular dependence. 
 In some parts of the phase diagram we encounter a divergence in the RG flow, indicating the onset of ordering with a gap that is in the detectable regime, i.e. larger than $10^{-3} t_f$. In other regimes, where such a divergence is not reached,  one can read off the dominant tendency of the RG flow.
In Fig.~\ref{Flow_HF} we show examples of RG flows as a function of
$\ell$. In Fig.~\ref{Flow_HF} (a), we show the competition between d-wave and
s-wave pairing, with d-wave being dominant and s-wave being
 subdominant. In Fig.~\ref{Flow_HF} (b) we show an example with dominant d-wave channel and subdominant  AF channel. In both cases we find that  for short distances (or high energies) CDW fluctuations are dominant, giving rise to a state that resembles the findings for high-$T_c$ superconductors. Note that in some situations the many-body states are almost degenerate and small changes in the initial conditions (that is, changes in the form of the interactions) can be used to select one particular ground state. 

With this procedure we determine
 the phase diagram of the system, which is shown in Fig.~\ref{PD}.
We  now discuss the general features of the phase diagram.
In the absence of any coupling to the bosons, i.e. for $\tilde{V}=0$, the
system shows s-wave pairing for attractive interaction, $U_{ff}<0$,
and no ordering for $U_{ff}>0$, i.e. Fermi liquid behavior, except for the special case of half-filling where Fermi surface nesting 
drives the system to AF order 
for repulsive 
interactions, and to s-wave pairing (degenerate with CDW) 
for attractive interaction. 
If we now turn on the interaction to the bosons, this picture is modified in
 the following way: The boundary of the s-wave regime is moved into the
 regime of positive $U_{ff}$, approximately to a value of $U_{ff}$ where the
 effective interaction at the nesting vector ${\bf Q}$ between the fermions,
 $U_{ff}+V_{{\rm ind.},\bold{Q}}$, is positive, i.e. for $U_{ff}\approx \tilde{V}/(1+8\xi^2)$. On the repulsive side, and
 away from half-filling,  we find the tendency to form a paired state, either
 d-wave or p-wave. This tendency becomes weaker the further the system is
 away from half-filling.  We typically find a gap in the vicinity of
 half-filling and further away from $\mu=0$ we find only an increasing strength of the corresponding interaction channel.  For the half-filled system, we find that for attractive interactions the degeneracy between s-wave pairing and CDW ordering is lifted, with s-wave pairing being the remaining type of order. For repulsive interactions, we find an intermediate regime of d-wave pairing, and for larger values of $U_{ff}$ we obtain AF order. 

\begin{figure}
\includegraphics[width=8.5cm]{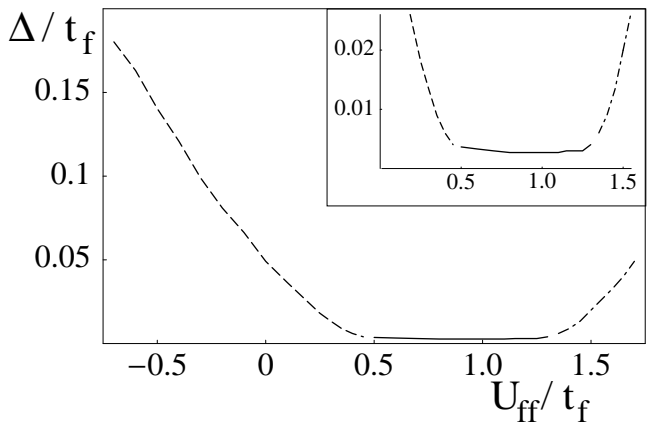}
\caption{\label{Gap_HF}
Many-body energy gaps at half-filling ($\mu=0$), as a function
 of $U_{ff}/t_f$, for $\tilde{V}/t_f=3$ and at a fixed value
 of the coherence length $\xi=1$. Dashed line: s-wave gap; Continuous line: d-wave gap; Dotted-dashed line: antiferromagnetic gap.  The inset shows a magnified plot of the d-wave regime.}
\end{figure}

The RG approach also allows the extraction of the many-body gaps
in the system through a "poor man's scaling" analysis of the divergent flow: at the point where the coupling becomes of order of $t_f$
the scaling parameter $\ell$ reaches the maximum value $\ell^* 
= \ln(t_f/\Delta)$, where $\Delta$ is the value of the gap. Hence,
$\Delta/t_f \approx \exp\{-\ell^*\}$ can be obtained from the RG
flows such as the ones in Fig.~\ref{Flow_HF}. In Fig.~\ref{Gap_HF} we show the gaps of the problem as a function of $U_{ff}/t_f$ in the half-filled case. One can see that as $U_{ff}$ increases, from negative to positive values, the s-wave gap is replaced by a d-wave gap, and finally for an antiferromagnetic gap. As is apparent from this figure, the gap in the d-wave phase is much smaller than the gaps of the AF order and the s-wave pairing, and, furthermore, almost independent
 of the value of $U_{ff}$. The latter is the case because the $U_{ff}$ term is a pure s-wave contribution to the interaction and therefore does not contribute to the d-wave  channel. The d-wave channel has an initial contribution which is entirely due to the anisotropy of the induced interaction, which gives only a small   value, and as a consequence only a small value for the gap. The value of the gap (in units of $t_f$) can be numerically fitted with a
 BCS expression of the form $a \exp(-b/\tilde{V})$, with the parameters $a$ and $b$ given by $a=0.31$ and $b=14.2$.

We have also performed RG calculations for a system of spinless fermions. This can be obtained by suppressing one of the spin indices in (\ref{Ham}) or (\ref{Ham_k}). In this case there is a major simplification in the problem since $U_{ff}$ is absent: in a spinless problem there can be only one fermion per site, as per Pauli's principle. Hence, in the absence of bosons, the spinless gas is non-interacting. The bosons, however, mediate the interaction between the fermions. Since the fermions are in different lattice sites the pair wavefunction has necessarily a node and hence, no s-wave pairing is allowed. In other words, in the spinless case the anti-symmetry of the wavefunction requires pairing in an odd angular momentum channel. In fact, we find that throughout the entire phase diagram the fermions develop $p$-wave pairing. At half-filling we find
a similar behavior of CDW fluctuations on short scales, analogous to the flow shown in Fig. \ref{Flow_HF}.  One should point out that in real solids the conditions of "spinlessness" behavior is hard to achieve since it usually requires complete polarization of the electron gas, that is, magnetic energies of the order of the Fermi energy (a situation experimentally difficult to achieve in good metals). However, in cold atom lattices this situation can be easily accomplished with the correct choice of atoms.

The many-body states discussed in this
 paper can be observed through various
 methods: AF order could be revealed in 
time-of-flight images and Bragg scattering \cite{Bragg}, 
 noise correlations \cite{noise} can be used to detect 
the various pairing phases,
 laser stirring experiments \cite{stirring} can be used
 to detect the phase boundary between AF order and pairing. 
  The short-scale CDW fluctuations should give a signature 
 in a photo-association measurement. RF spectroscopy \cite{chin} can be used
 to quantify the gaps of the various phases.

In summary, we have used a functional RG approach to study BFM of ultra-cold atoms in a 2D optical lattice. We found a number of competing phases, in particular in the vicinity  of half-filling, including AF ordering, s-, p-, and d-wave pairing. 
Our approach enables us to  quantify the gaps of these phases, to identify subdominant orders, and to study short-range fluctuations. Optical lattices of cold atoms allow for a unique opportunity of study of complex many-body states under well controlled circumstances, a situation that can be hardly found in real solids.

We thank E.~Altman, E.~Demler, and A.~Polkovnikov for illuminating conversations. 
A.~H.~C.~N. was supported by the NSF grant DMR-0343790.



\begin{thebibliography}{99}
\bibitem{remark_spin}
$U(\bold{k}_1,\bold{k}_2,\bold{k}_3)$ denotes processes
involving electrons with opposite spins. Processe involving parallel spins 
can be obtained from these \cite{Zanchi_Schulz}.
\end{thebibliography}
\end{document}